# REAL-TIME SYSTEM FOR HUMAN ACTIVITY ANALYSIS

*Randy Tan* *Naimul Khan* *Ling Guan*

Department of Electrical and Computer Engineering, Ryerson University

## ABSTRACT

We propose a real-time human activity analysis system, where a user's activity can be quantitatively evaluated with respect to a ground truth recording. We use two Kinects to solve the problem of self-occlusion through extracting optimal joint positions using Singular Value Decomposition (SVD) and Sequential Quadratic Programming (SQP). Incremental Dynamic Time Warping (IDTW) is used to compare the user and expert (ground truth) to quantitatively score the user's performance. Furthermore, the user's performance is displayed through a visual feedback system, where colors on the skeleton represent the user's score. Our experiments use a motion capture suit as ground truth to compare our dual Kinect setup to a single Kinect. We also show that with our visual feedback method, users gain statistically significant boost to learning as opposed to watching a simple video.

## I. INTRODUCTION

Human activity analysis is the process of automated evaluation of different actions. Activity analysis has a wide range of applications including: physical rehabilitation, assisted living, telemedicine, entertainment, and fitness. In many of these application areas, people wish to learn how to perform certain activities. Traditionally, a user learns through repetition while an expert observes and provides qualitative feedback. Some problems with this approach is that a human expert has to be present, and the feedback is mostly qualitative so it varies between experts. A universal automated activity analysis process can greatly enhance the quality of life for many people, especially when the expert is not present for the repetitive part of learning.

Early research in making activity analysis systems used elaborate motion capture systems which were expensive, thus reducing accessibility. RGB-D devices are a cheap and commercially available alternative. Additionally Microsoft's Kinect Software Development Kit (SDK) offers real-time skeletal tracking. There have been a few studies on applications for Kinect in training applications [1], [2]. Unfortunately, these methods only classify exercises performed as either *correct* or *incorrect* and only at the end of the exercise. A more recent work using Kinect provides real-time evaluation of exercises in the form of qualitative audible feedback [3]. While their system had positive feedback from users, their quantitative results were inconclusive. To provide meaningful evaluation of an exercise, the system needs to be able to isolate the source of errors.

While the Kinect offers a cost effective method of depth sensing, errors caused by occlusion limits natural movements. Several methods have been proposed to correct failed tracking [4], [5]. In [4], a computationally efficient method is proposed where the skeletons from dual Kinects, facing the user at different perspectives, are used to synthesize a new skeleton. We improve upon their system with better initialization to prevent the optimization algorithm from getting stuck at a local maxima (details in Section II-A). While there are more accurate methods such as [5], and even commercial systems such as ipisoft (http://ipisoft.com/), the accuracy comes at the price of high computational costs, prohibiting their usage in real-time.

Our system which was initially proposed in [6], [7] was designed to allow users to intuitively gauge their performance in real-time. Incremental Dynamic Time Warping (IDTW) was used to match the user to a pre-recorded ground truth as it allows variation in speed. The IDTW costs were calculated per limb and displayed back to the user with a color-coded skeleton visualization, so that they can identify which limbs are the source of error if they do not perform the action properly. Using a color-coded image as a visual measure of performance transfers information to the spacial portion of working memory better than printed text [8]. Also, any audio cues that convey qualitative feedback can still be used because they use verbal working memory. While the system showed potential in evaluating human activities, we solve several of their limitations including:

* **The limited number of natural movements with only a single Kinect.** One of the most serious weaknesses of vision-based sensors is their dependance on perspective. The use of multiple sensors increases the range of natural movements that could be recorded as long as at least one Kinect can see each body part.
* **The reliance for the user and expert to use the same Kinect setup.** The goal of our system is to allow users to practice activities unsupervised by an expert. Our improved calibration stage allows Kinects to be moved

around in between the expert session and the user session.

* **The lack of quantitative experimentation to prove the effectiveness of the system.** Our previous works [6], [7] had limited user experimentation. [6] only demonstrated the visual feedback without user experiments and [7] did not have a control group. The experiments done in the original dual Kinect work measured the variation in limb length, which discarded important orientation information, and did not reference a ground truth [4]. Additionally the comparison done between single, and dual Kinects, did not place the single Kinect in its optimal viewing position (front facing). Our experiments show a statistical improvement in user learning rate when compared to trials with visual feedback disabled. Additionally they show that our dual Kinects setup increases the range of trackable movements; using a motion capture suit as ground truth.

## II. THE PROPOSED SYSTEM

### II-A. Skeletal Voting

Two Kinects are used in a voting system to solve the problem of self-occlusion as adapted from [4]. When our system is run, calibration can be initiated by the user clicking a button when they see both Kinects are tracking properly. This calibration is used to find the rotation and translation between the two Kinects. We designate an arbitrary Kinect as the main Kinect and all other Kinects are transformed to this space; using rigid body transform [9] achieved via SVD. Our current setup uses two Kinects, but the voting system is based on minimizing a weighted average and can therefore be extended to multiple Kinects. As long as at least one Kinect can see all joints at any given time, more Kinects do not provide significant improvement. Using the calibration data the skeletons are tranformed and averaged, to obtain the user's initial rotation based on the vector from left hip to right hip, and the limb lengths.

The skeleton voting system uses SQP to optimize finding joint positions. For each joint, weights are used to identify which Kinect has the more accurate reported position. Let, $A$ and $B$ denote the two Kinects, and $S_A$ and $S_B$ represent the skeletons produced by them respectively. Each joint $i$ has positions $P_{A_i}$ $i \in S_A$, and $P_{B_i}$ $i \in S_B$ reported by each Kinect. Our target is to calculate the weights $w_{A_i}$ and $w_{B_i}$ for each joint representing how reliable the reported positions are. Kinect reports mistracked joints as either *inferred* or the reported position is a foreground pixel which is not the correct joint position. As long as there are no objects occluding the user, any incorrect positions reported as *tracked* are from self-occlusion and therefore close to another reported unconnected joint. Hence, the first step to calculating the weights is to calculate the minimum distance between each tracked joint and the closest unconnected joint:

$$d_{A_i} = \min_{k \in S_A, k \neq i,c} ||P_{A_i} - P_{A_k}||^2 \qquad d_{B_i} = \min_{k \in S_B, k \neq i,c} ||P_{B_i} - P_{B_k}||^2 \tag{1}$$

Where $d$ is the distance of joint $i$ to the closest reported joint $k$ in skeleton $S$ except for itself and any connected joint $c$. The distances are normalized to form the initial part of the weighting as shown as:

$$\bar{w}_{A_i} = \frac{d_{A_i}}{d_{A_i} + d_{B_i}} \qquad \bar{w}_{B_i} = \frac{d_{B_i}}{d_{A_i} + d_{B_i}} \tag{2}$$

To incorporate the Kinect's tracking state into the weighting, variables $h_{A_i}$ and $h_{B_i}$ are used in creating the final weights. Their value will be assigned a value of $\theta$ if that Kinect (A or B respectively) is tracking the joint $i$ and set to $(1-\theta)$ if the joint position is being inferred where $\theta \in (0.5, 1)$ and can be tuned. All examples in this paper use $\theta = 0.9$. By using a larger $\theta$, the weighting relies more on the joint being reported as tracked. The final weights are normalized again as shown by by:

$$w_{A_i} = \frac{(\bar{w}_{A_i} h_{A_i})^4}{(\bar{w}_{A_i} h_{A_i})^4 + (\bar{w}_{B_i} h_{B_i})^4} \quad w_{B_i} = \frac{(\bar{w}_{B_i} h_{B_i})^4}{(\bar{w}_{B_i} h_{A_i})^4 + (\bar{w}_{B_i} h_{B_i})^4} \tag{3}$$

By adding the power of 4 to the normalization scheme, the differences between weights are heavily exaggerated. If one of the pre-normalized weights were significantly higher than the other, after normalization with the power of 4 the lower weight would be pushed down to almost 0 while the higher weight would be pushed up to almost 1.

After weighting the reliability of the reported joint positions, Sequential Quadratic Programming (SQP) is used to find the final optimal joint positions. SQP takes a quadratic optimization function and a set of quadratic constraint functions and iteratively finds a solution using Newton's Method. The objective function used in [4] is to minimize the weighted sum of distances between the final voted joint position and the reported position of each Kinect while the constraint is that the limb length found during the calibration stage must be preserved. The Lagrange function for the described problem can be shown as:

$$\begin{aligned} \ell(P_{V_i}, \lambda) = &\sum_{i \in S_A, S_B} w_{A_i} ||P_{V_i} - P_{A_i}||^2 + w_{B_i} ||P_{V_i} - P_{B_i}||^2 \\ &+ \lambda \sum_{i,j \in S_A, S_B} ||P_{V_i} - P_{V_j}||^2 - l_{i,j}^2 \end{aligned} \tag{4}$$

where $j$ is the parent joint which is the next connected joint to $i$ leading to the center hip, $l_{ij}$ is the limb length that was recorded in the calibration stage, and $\lambda$ is the Lagrange multiplier.

A visualization of the voting can be seen in 1. In [4], the SQP was initialized at voted position from the previous frame. In our system, we set our initial guess for each joint as the position of the Kinect with the higher weighting. Through testing, we found that it was possible with the old initialization to get stuck at a local maxima. This is due to the fact that SQP only looks for the closest local optimal point which can be a max or min and is therefore sensitive to initialization. Looking at figure 1, The final joint position needs to be on the sphere with the radius defined by the limb length. On that sphere, the minimum point should be located closest to where the reported joint positions with high weights are and the maximum point is on the opposite side. If the previous voted joint position is closer to the max point due to either fast movements or accidental mistracking, then under the old initialization the joint will get voted to the max point and the initialization for the next frame's voting will always be closer to the maximum. Under the new initialization, as long as the Kinect with the higher weight is tracking properly, the initialization will always be closer to the min point. We end the line search when the voted point moves less than 0.1*mm* or when 50 iterations have passed.

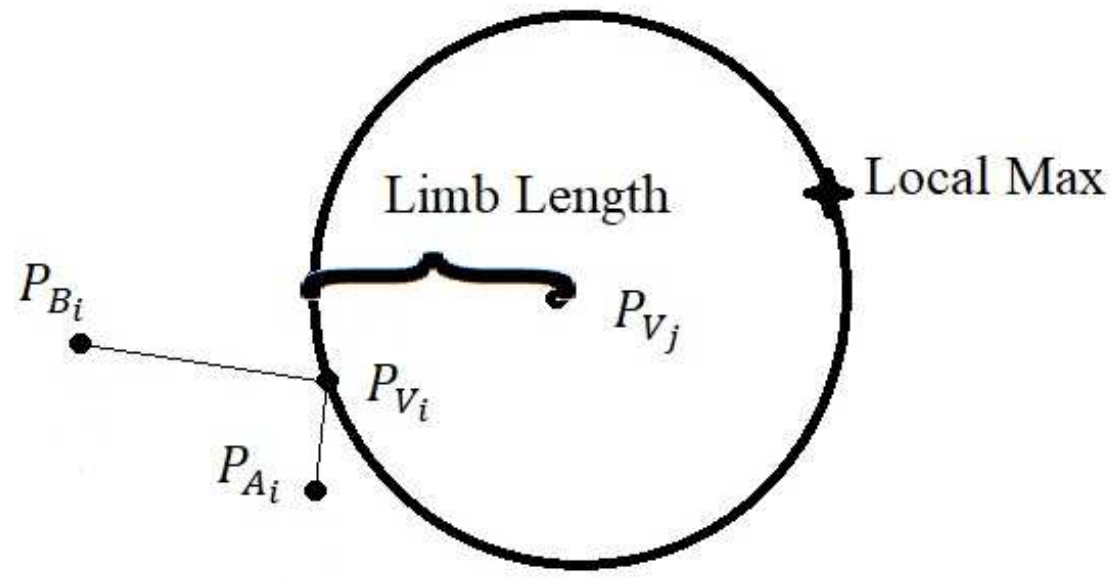

**Fig. 1**: **Voting for one incorrect joint**. $P_{A_i}$(tracking) and $P_{B_i}$(mistracking) are the reported joint positions from Kinect A and B. The radius of the circle represents the limb length obtained in calibration. With $P_{V_i}$ being the final voted joint, restricted by limb length.

### II-B. The IDTW Algorithm and Grading

We use a similar grading scheme that was proposed in [6]. Two major improvements are:

* We removed the requirement that for a limb to be graded, there has to be a minimum amount of movement of that limb. While this feature made the algorithm computationally more efficient, certain exercises require proper posture meaning that a limb can not be disregarded from grading just because it does not move.

* During the calibration stage for multiple Kinects, the vector between the left hip and right hip is saved and every skeleton is rotated by that vector before being sent to the grading algorithm. This allows our grading system to be calibration dependent rather than Kinect placement dependent. As a system that is meant to be used without the expert being present, we must assume that the Kinect setups will not have identical placement.

Grading using IDTW is done on a per-joint basis. Each joint's position is normalized by having the coordinates from the parent subtracted and the result is divided by limb length. Each normalized joint coordinate can be calculated by $J_i = \frac{P_{V_i} - P_{V_j}}{||P_{V_i} - P_{V_j}||}$, where $P_{V_i}$ and $P_{V_j}$ represent the voted joint locations as stated before. This normalization allows the system to accommodate for people with different limb sizes [10].

Time sequences of the user $U$ with N frames and the expert $E$ with M frames are compared in a grid. For each cell ($U_a$,$E_b$) in the grid, the distance between the normalized joint positions $J_i$ are compared for the sequence $U$ at time frame $a$ and $E$ at time frame $b$. The equation for IDTW is [6]:

$$\mathbf{D}_i(\mathbf{U},\mathbf{E}) = \min_{c=1,\ldots,M} \frac{1}{N} \sum_{t=1}^{T} ||{J_i}^{U_t} - {J_i}^{E_t}||, \qquad i \in S_U, S_E. \quad (5)$$

Where T is the total grid cells taken in the warping path, t corresponds to each grid coordinate $(U_a, E_b)$ in that path, and $S_U$ and $S_E$ are the set of all joints in the skeletons U and E. In the classical DTW approach, the minimum path is required to reach from the bottom left of the grid to the top right. Since our application is real-time, the full sequence for the user is not complete and therefore the requirement for the sequence to reach the top right needs to be relaxed. IDTW achieves this by requiring the warping path to include all frames of the user sequence, but only using the first $c$ frames of the expert sequence that minimizes the DTW cost. Finally, the limb scores are calculated as:

$$Z_{i,j} = e^{-\nu*(\mathbf{D}_i+\mathbf{D}_j)/2}, \quad (6)$$

Where $i$ and $j$ are the joints that form this particular limb $Z_{i,j}$, and $\nu$ is a parameter to control the score's sensitivity. In our experiments, $\nu$ was set to 10 for all joints above the hips and 30 for the hips and below as we found that the legs had an easier time keeping a good score. The $Z_{i,j}$ scores are projected onto the skeleton visualization through a color map to form the visual feedback system. The color map used in this work is blue-aqua-green-yellow-red, i.e., the color stays closer to blue if the user is doing well, and shifts towards red as the performance worsens.

## III. EXPERIMENTAL RESULTS

The goal of these experiments is to verify that the algorithms can track and evaluate users and show that using the system can improve performance. To our knowledge, this is the most comprehensive set of experiments that objectively demonstrate the performance of Kinect based activity analysis. However, direct comparison to other methods is not possible due to the fact that experiments for activity analysis systems are based on how the systems were designed. Our system is meant to teach any kind of activity while other systems may target teaching a specific activity or increasing the user's fitness. Many works such as [11], [3], [1] have mixed results for user testing, failing to convincingly demonstrate their claimed advantages. For our experimental setup, our system is implemented using a Visual Studio WPF application. The computer hardware was an i7-4790 at 3.6GHz with 16GB RAM. The system ran between 3 to 4 ms per frame on average with both skeleton voting and IDTW running which is in-line with [4]. The experiments are separated into three sections.

The first part of the experiment shows the system's improved tracking compared to a single front facing Kinect using a XSens motion capture suit as ground truth (https://www.xsens.com/products/xsens-mvn/. The second part of the experiment shows that our system can differentiate between good and poor performances pinpointing the source of errors. The third part of the experiment shows that our real-time feedback system can allow users to learn an activity much more efficiently than simply watching and imitating a video. In all of our tests, we had 2 lab members act as the *expert* and 8 participants act as *users* with a mix of different gender, age, and demography. In all parts of the experiment, four simple exercises (bar curl, horse stance (from karate), marching, vertical press) and two complicated Tai Chi exercises (brush Knee, parting the wild horse's mane) are used. The scores shown are the IDTW costs totaled across all joints; lower scores are better. While the numeric scores can give an indication of overall performance, it does not give a clear picture of when and where mistakes occurred. For clarifications readers are recommended to watch the video demonstrations of the exercises used in the experiment at: https://youtu.be/GMkbgja5Nng and https://youtu.be/zp1BjMsK22M.

## III-A. Multi-Kinect Performance Test

In the tracking test, only the ground truth actors are used for comparing Kinects to motion capture suits for all 6 exercises. The first actor performed *bar curl, marching, and brush knee* while the other actor performed *horse stance, vertical press, and horse's mane*. The Xsens MVN mocap suit is used as ground truth since it is an inertial based motion capture system meaning that it can not be self-occluded. The actors performed each exercise with the dual Kinects and single Kinect separately while wearing the MVN suit for ground truth. The skeletons were synchronized at 30 FPS and a rigid body transform was used to match their coordinate systems. Joint positions were normalized relative to parent exactly like how our system process features. The exercises were recorded multiple times to make sure the mistracking consistently occurred for the single Kinect; only one session is shown. The figures 2 and 3 shown are for *horse's mane* and *brush knee* since they highlight obvious mistracking for single Kinects. The graphs show the distance per frame between features created by the Kinect(s) and Xsens. The motion capture suits places joints in slightly different locations than the Kinect so there will always be a static amount of distance between them. Areas of interest in the graphs will therefore be the spikes in the graphs which are marked for single Kinect. For every single Kinect error frame, we show the closest corrected Dual Kinect frame. For every marked frame, the 3D skeletons are shown from a perspective that shows the errors best. As seen in both of these figures, a single front facing Kinect loses tracking in both of the Tai Chi exercises while the dual Kinects maintain tracking.

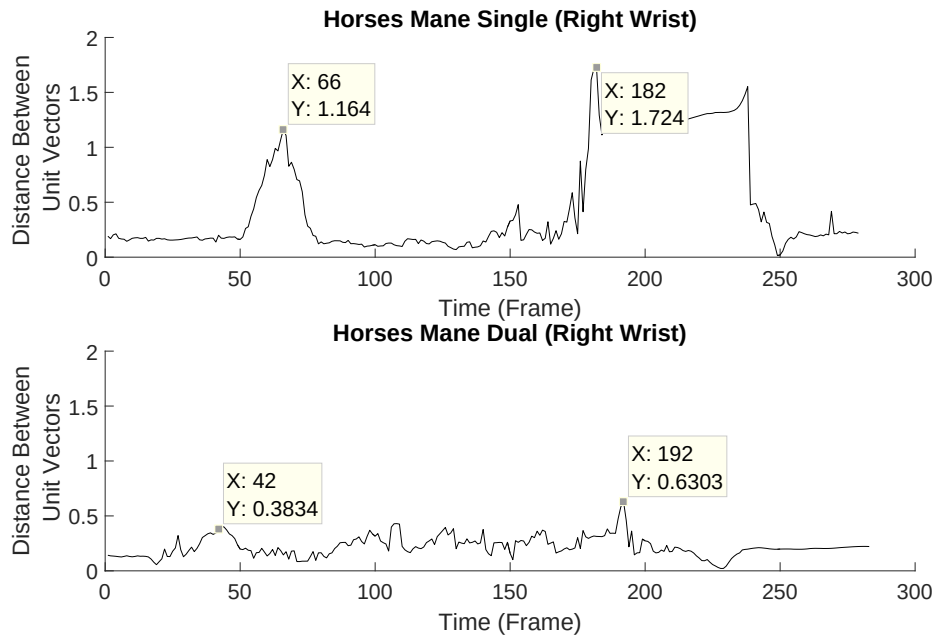

(a) Distance scores for right wrist in horse's mane. Top graph is the single Kinect, the bottom graph is the dual Kinect.

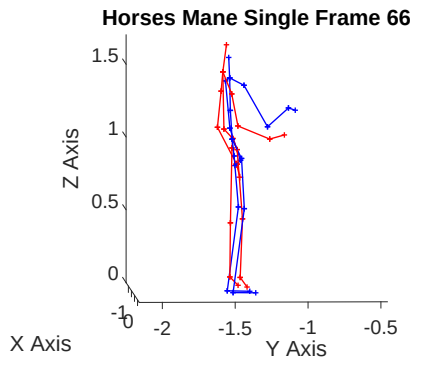

(b) Single Kinect: delay in raising wrist

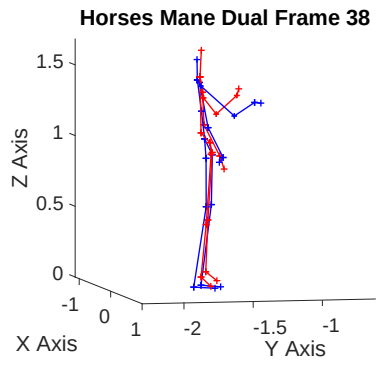

(c) Dual Kinects: no delay in raising wrist

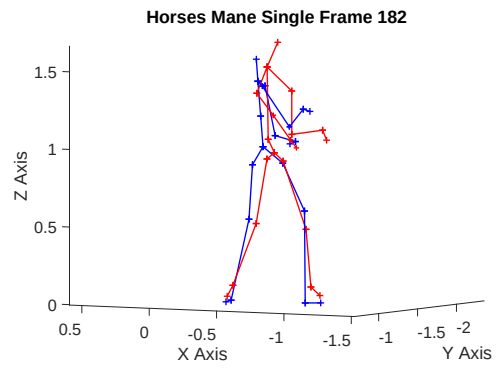

(d) Single Kinect: right wrist mistracking

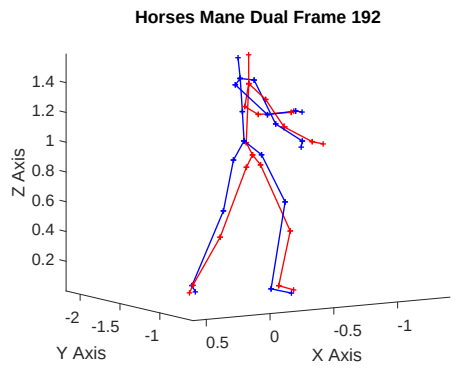

(e) Dual Kinects: right wrist tracking properly

**Fig. 2**: Tracking results for horse's mane exercise. The spikes marked in (a) for single Kinect are shown in (b) and (d) while equivalent dual Kinect frames are shown in (c) and (e). Blue skeletons belong to the motion capture suit and red skeletons belong to the Kinect

## III-B. Grading Test

The grading test includes the four simple exercises in which each participant knew the exercises ahead of time and was asked to perform the exercise properly five times and

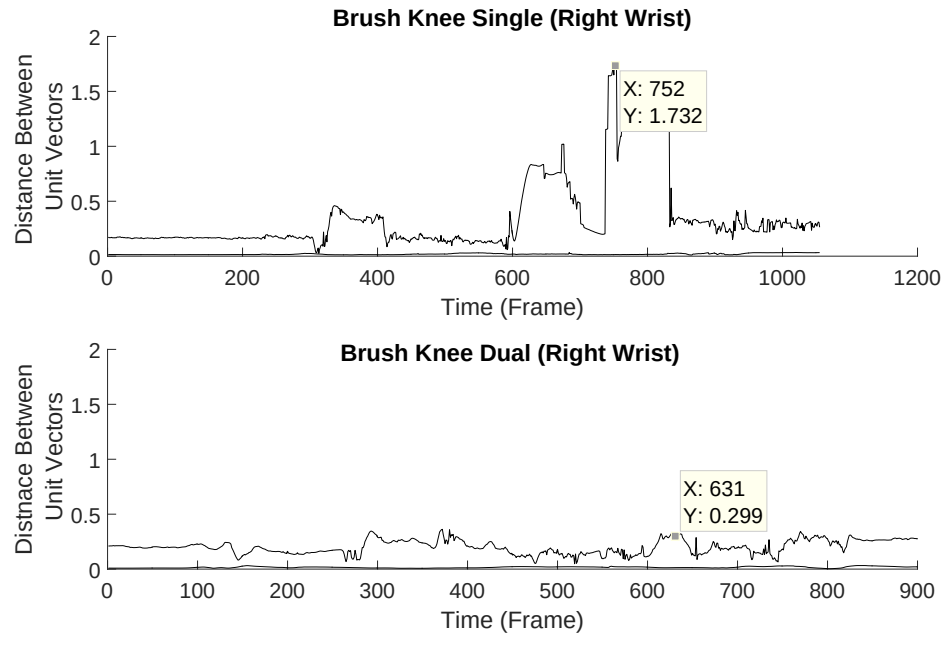

(a) Distance scores for right wrist in brush knee

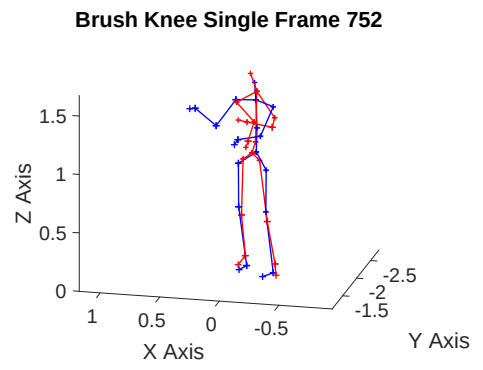

(b) Single Kinect: right wrist mistracking

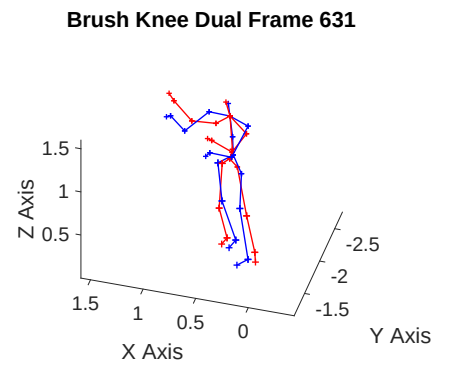

(c) Dual Kinects: right wrist tracking properly

**Fig. 3**: Tracking results for brush knee exercise. The spike marked in (a) for single Kinect is shown in (b) while the equivalent dual Kinect frame is shown in (c). Blue skeletons belong to the motion capture suit and red skeletons belong to the Kinect

then make a specific mistake five times. The exercises and mistakes were: a) vertical press with inclined back, b) march with not bending the legs 90°, c) bar curl with putting the whole arm into motion instead of only using biceps and d) horse stance (from karate) with not spreading the legs far out enough and compensating by pointing the knees outwards.

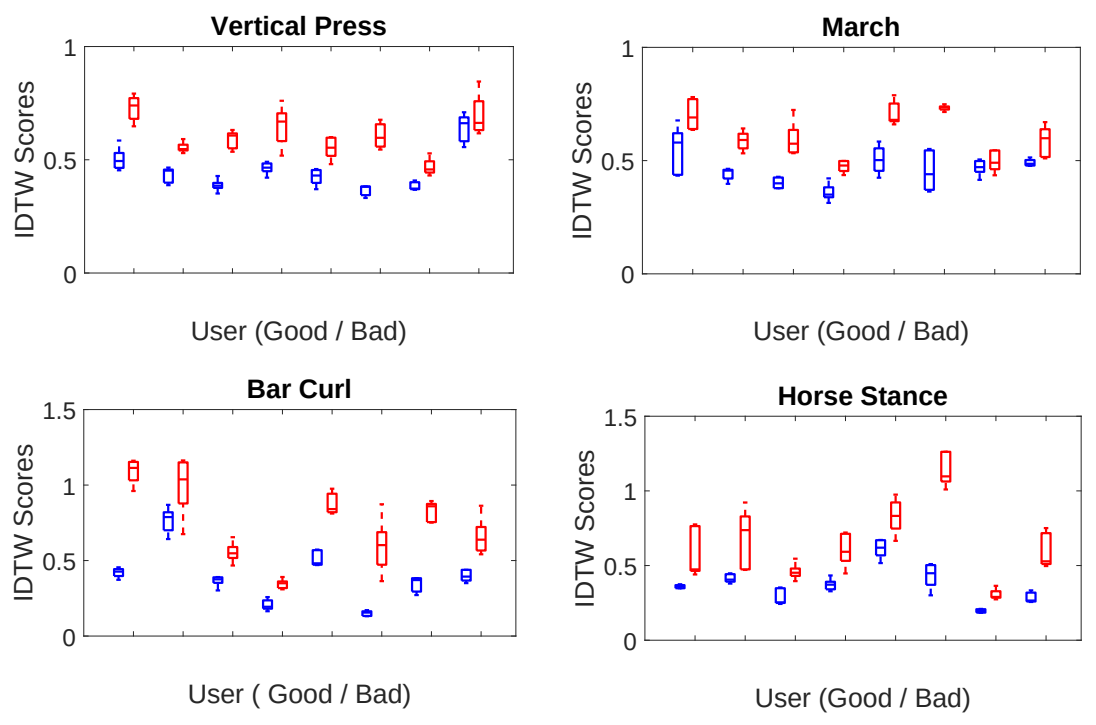

**Fig. 4**: Results of the grading test. Each user's good performances shown in blue, while incorrect shown in red. The results are shown in a boxplot representing the variance for each user over five sessions.

The results of the grading test in Figure 4 show the effectiveness of IDTW scoring. For each individual user, the incorrect performances received higher IDTW costs on average compared to their proper performances. It is also important to note that the IDTW costs were fairly dependent on the user and the type of exercise. While it may not be possible to set a universal threshold between good and poor performances, it is possible to easily tune the visual feedback per individual and exercise by changing $v$ in equation 6.

### III-C. User Study

The objective of the user study is to quantify a user's performance over consecutive sessions with/without the colored skeleton visualization while following a difficult routine alongside an expert. We chose two Tai Chi exercises *Parting the Wild Horse's Mane* and *Brush Knee*. Tai Chi was chosen as it was complex enough to require practice while not being strenuous to perform. In order for the users to learn the exercises quicker, the ground truth recordings break the exercises up into individual steps with pauses in between (shown in the videos). The eight participants were divided into two groups: the first group performs Horse's Mane with the visual feedback system enabled while performing Brush Knee with the visual feedback system disabled and vice versa for the second group. All of the users confirmed they did not know how to perform the exercise before the experiment. Each user performed each exercise for ten sessions. The users' cumulative IDTW costs were recorded at the end of each session.

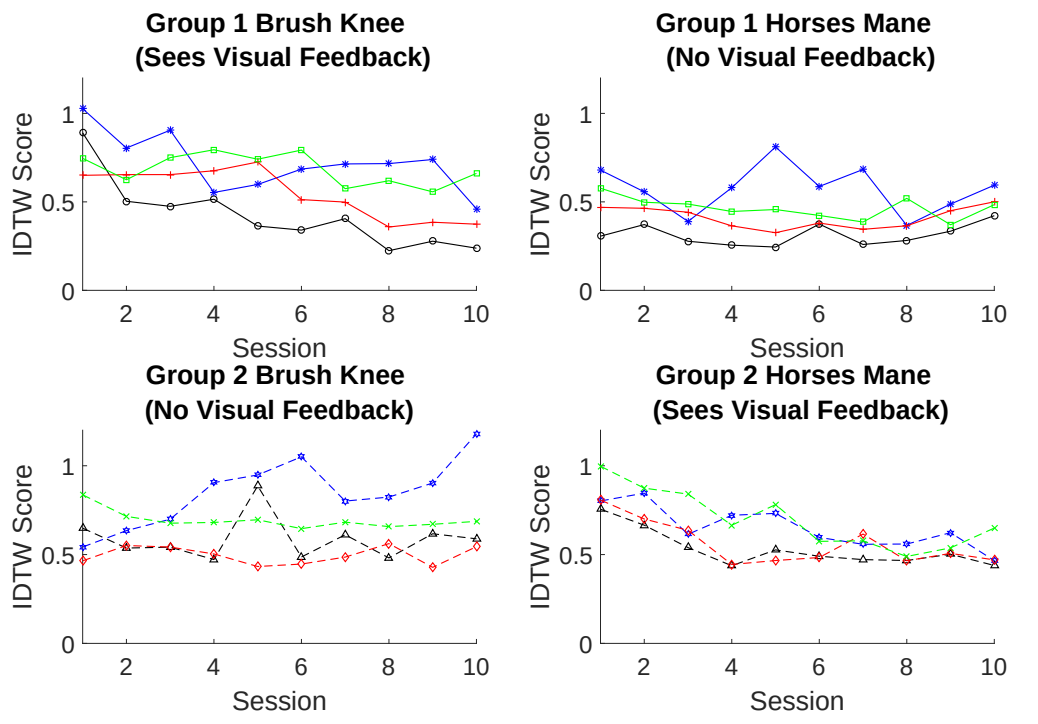

**Fig. 5**: Results of the user study. Each graph shows each group's individual DTW costs over 10 sessions for each of the two exercises

As seen in Figure 5, when each group saw their feedback there was a downward trend in their IDTW costs. When the colored skeletons were disabled, the user's IDTW costs were more erratic overall and some users performed worse over the ten sessions. While it can be argued that certain users performed better than others and that both of the exercises were not equal in difficulty, erratic scores between sessions only occurred when users did not see their visual feedback. To prove that users improved over 10 sessions when they saw feedback, we preformed statistical trend analysis. After all the sessions were recorded, the samples were grouped by *sessions with visual feedback* and *sessions without visual feedback*. A Mann-Kendall test was run on the median per session of each group [12]. The purpose of the Mann-Kendall test is to identify if there are any monotonic trends in a series of data. The median was recommended by [12] for central values with small sample sizes. The hypothesis of the test is that there are no trends in the data. Typically, a p-value of less than 0.05 means that the hypothesis can be rejected with confidence indicating there is a trend in the series. Table I shows the median samples per session and the results of the Mann-Kendall test. The sessions with feedback received a p-value of 0.0013 meaning that there is a trend present in the data while the group without feedback failed to reject the hypothesis. These results demonstrate that the visual feedback system can indeed help a user quickly improve over time with easy-to-interpret feedback, which is not possible with simply imitating a video.

**Table I**: Results of Mann-Kendall trend test on the median IDTW of all users *with feedback (WFB)* and *no feedback (NFB)*

| Session | 1 | 2 | 3 | 4 | 5 | 6 | 7 | 8 | 9 | 10 | P-value |
|---|---|---|---|---|---|---|---|---|---|---|---|
| WFB | 0.8043 | 0.6828 | 0.6454 | 0.6092 | 0.6618 | 0.5436 | 0.5662 | 0.4780 | 0.5222 | 0.4635 | **0.0013** |
| NFB | 0.5598 | 0.5438 | 0.5147 | 0.4877 | 0.5763 | 0.4665 | 0.5486 | 0.5005 | 0.4688 | 0.5666 | 0.5195 |

## IV. CONCLUSION

In this paper, we propose an improved real-time human activity analysis system. Our system improves the range of trackable movements through the use of dual Kinect voting using SVD and SQP. IDTW is used to calculate the distance cost between a partial user sequence and a complete expert sequence. Visual feedback allows users to easily understand where they can improve their performance in real-time. Improvements were made to the system to allow different Kinect placements between sessions. Our experiments demonstrate the increase range of movements of the dual Kinect setup without self-occlusion. Our experiments also show a statistically significant learning trend when users saw our visual feedback which did not occur when the visual feedback was disabled. Our activity analysis system can be used to improve learning for users while they practice activities during the times when an expert is not present.